\documentclass[A4,twocolumn,epsf,nofootinbib,showpacs,prd]{revtex4}
\usepackage{amssymb}
\usepackage{epsfig}
\usepackage{graphicx}
\usepackage{psfrag}
\usepackage{amsmath,pstricks}

\DeclareMathAlphabet   {\mathsc}{OT1}{cmr}{m}{sc}  
\def\[{\left [} 

\def\]{\right ]} 

\def\({\left (} 

\def\){\right )}

\newcommand{\GeV}      {~\mathrm{GeV}}

\def\be{\begin{equation}}

\def\ee{\end{equation}}

\newcommand{\gappeq}{\mathrel{\rlap {\raise.5ex\hbox{$>$}} 

{\lower.5ex\hbox{$\sim$}}}} 

\newcommand{\lappeq}{\mathrel{\rlap{\raise.5ex\hbox{$<$}} 

{\lower.5ex\hbox{$\sim$}}}}

\newrgbcolor{darkred}{0.7 0 0}

\newrgbcolor{green}{0 1 0}

\newrgbcolor{darkblue}{0 0 0.7}

 \newrgbcolor{purple}{1 0 1}

\begin{document}

\title{Halo Geometry and Dark Matter Annihilation Signal}
\author{E. Athanassoula$^a$, F.-S. Ling$^b$, E. Nezri$^b$}
\affiliation{$^a$ Observatoire de Marseille, 2 Place Le Verrier, 13248 Marseille C\'edex 04, France\\
$^b$ Universit\'e Libre de Bruxelles, Campus de la Plaine CP225, Belgium
}

\vspace{0.5truecm}

\begin{abstract}

We study the impact of the halo shape and geometry on the
expected weakly interacting massive particle (WIMP) dark matter annihilation signal from the galactic center.
As the halo profile in the innermost region is still poorly constrained, 
we focus on geometrical distortions and consider different density behaviors 
like flat cores, cusps and spikes.
We show that asphericity has a strong impact on the annihilation signal 
when the halo profile near the galactic center is flat, but becomes
gradually less significant for cuspy profiles, and negligible in the
presence of a central spike. However, the astrophysical factor 
is strongly dependent on the WIMP mass and annihilation cross-section
in the latter case.
\end{abstract}


\maketitle

\vspace{1truecm}

\section{Introduction}
Flat rotation curves of spiral galaxies \cite{bosma0} can be
explained by the presence of a dark matter halo which extends much
farther than the  
luminous disc. While at large distances the gravitational potential is
completely dominated by the dark halo, there is still a vivacious debate about
whether the dark matter is prevailing in the central parts of 
bright galaxies, and about whether its radial matter
distribution is cuspy, or not~\cite{bosma1}. 
Furthermore, very little is known with certainty about the shape of
the halo of disc galaxies and in  
many cases the halo is simply assumed to be 
spherical. 

However, if the disc is indeed an important component in the central
parts, it should, due to its gravity, introduce some flattening of 
the dark matter distribution.  Furthermore, large scale cosmological
$N$-body simulations have shown  
that, at least at large distances from the center, the natural shape of
dark halos is triaxial (see references in \cite{natara}) with density
axial ratios in the range 0.5 -- 0.8 (\cite{Jing}). It is thus  
natural to ask whether the flatness and the various complex 
structures and substructures of the luminous part of the galaxy will affect
the dark matter halo, and 
to what extent the triaxiality of the halo will change a possible dark
matter annihilation signal from the central parts of a galaxy.

Usually, two types of asphericity are considered : 
flattening of the halo and departure from axisymmetry. 
The halo flattening is quantified by the value of $q=c/a$,
where $a$ is the major axis in the galactic
plane and $c$ is the axis perpendicular to that plane.
Various observational methods have been used to probe
the halo flattening in our own Galaxy and in neighboring
ones~(e.g. \cite{natara,bosma2}). It was found that
the measured flattening can vary over a wide range of values, 
depending on the galaxy and on the method used. A cross-check of the
different methods  
with their systematic biases on the same galaxy would be welcome but is 
usually not possible. Measurements based on atomic hydrogen favour
oblate halos with shortest-to-longest ratios in the very wide range
of 0.2 to 0.8 \cite{merri}. For our own galaxy and based on the
thickness of the Milky Way's gas layer, ~\cite{om} argue for 
a rather round halo with flattening $0.8 \leq q \leq 1$, but their result
depends heavily on the values of the galactocentric radius $R_0$ and
of the galactic rotation speed $v_0$~\cite{om}. Recent studies of the
dynamics of the stellar stream coming from the disruption of the
Sagittarius dwarf galaxy also give a wide range of values, between 0.5
and 1.7~\cite{sagi}. 

The second type of asphericity is a departure from axisymmetry in the
galactic plane. This is statistically quite common as a large
fraction (more than 70\%) of present day disc galaxies have bars or 
ovals~\cite{EG}. Also, it is now well agreed that our Galaxy is
barred in its central parts~\cite{Dwek}. Bars form naturally also in
$N$-body simulations, as witnessed already in the early
seventies~\cite{MH}. More recently, it was realized 
that the presence of a dark halo can play an active role in the formation 
of the
disc bar, if it is non rigid, {\it i.e.} if it can interact with the disc. 
Indeed, bars evolve and grow stronger by the redistribution of angular
momentum within their 
galaxy. This is emitted by near-resonant material at the inner disc and
absorbed by near-resonant material in the outer disc and in the
halo~\cite{ath1}. As a result, the halo also is deformed and acquires
a bar structure, which is fatter and shorter than the disc bar, but
can concern a considerable amount of mass~\cite{ath2}.  

On the observational side, departures from axisymmetry can be checked from
the orbits of the baryons, in particular the HI gas that has low
velocity dispersion. Obtaining a quantitative estimate of such
asymmetries is, however, not trivial, since it implies a decoupling of
the halo contribution from that of the luminous matter as well as a 
knowledge of the inclination of the galactic disc~\cite{MR}. 
Of course, a direct probe of the halo would enable to see whether the halo
deformation follows the barred structure of the disk or not.
If the recent EGRET diffuse gamma ray signal above 1 $\GeV$ is interpreted as 
originating from dark matter, it indeed leads to such a structure with
an ellipticity value $0.65 \pm 0.15$~\cite{dB}.

The purpose of this article is to study the impact of an elliptical
deformation of the halo on the expected weakly interacting massive particle (WIMP) dark matter annihilation
signal from the galactic center. However, as discussed above, we do
not know for certain what the dark matter radial profile is, so that
distributions with or without a cusp, and with or without a spike have
to be considered. The density enhancement in a cuspy profile follows
the deepening of the central gravitational  
potential due to the baryon cooling through radiative processes
\cite{Blumenthal:1985qy,Edsjo:2004pf,Prada:2004pi,Gnedin:2004cx}. 
The presence of a super-massive black hole (SBH) at the galactic center
can further create a spike, or an enhancement of the cusp, at very short distances from the galactic center \cite{silkgondo,zhao}
but scatterings on stars and capture of dark matter particles by the SBH could 
decrease the density in this region \cite{gnedin,merrit04}.
It is clear that the presence of a spike would boost the annihilation signal by several 
orders of magnitude. 
In the sequel, we will see how elliptical distortions interplay with density profiles
in the dark matter annihilation signal.

\section{Halo parameterization and flux calculation}

Since the main observable annihilation signal from the galactic center is with 
$\gamma$-rays \cite{mansilk}, we will restrict ourselves to this case.
The observed gamma-ray flux of energy $E$, from the 
annihilation of dark matter particles $\chi$ (with mass $m_\chi$ and density $\rho$) 
and annihilation cross section  $\sigma_i v$ (into final state $i$), 
can be expressed as (e.g.~Ref.~\cite{JKG})
\begin{equation}
\frac{\Phi_{\gamma}}{d \Omega d E}= \sum_i \frac{1}{2}
\frac{dN^i_{\gamma}}{dE_{\gamma}} \langle \sigma_i v \rangle \frac{1}{4 \pi m_\chi^2}
\int_{\mbox{l.o.s.}}
\rho^2 \,\, d l ,
\label{flux}
\end{equation}
where $dN^i_{\gamma}/dE_{\gamma}$ is the differential gamma spectrum  per
annihilation coming from the decay of annihilation products of final state $i$ and
the integral is taken along the line of sight.
It is customary (see~\cite{BUB}), in order to separate 
the factors depending on astrophysics from those depending only 
on particle physics, to introduce the quantity $J(\vec{\Omega})$
for the line of sight $l(\vec{\Omega})$ corresponding to the direction
$\vec{\Omega}$,
\begin{equation}
J(\vec{\Omega}) = \frac{1} {8.5\, \rm{kpc}} 
\left(\frac{1}{0.3\, \mbox{\small{GeV/cm}}^3}\right)^2
\int_{l(\vec{\Omega})}\,\rho^2\,\, d l\,.
\label{gei}
\end{equation}
We then define the astrophysical factor 
$\bar{J}(\Delta\Omega)$ as  the average of $J(\vec{\Omega})$ over
a spherical region of solid angle $\Delta\Omega$, centered on the direction 
of the galactic center
\begin{equation}
\bar{J}(\Delta \Omega)= \frac{1}{\Delta \Omega} \int_{\Delta \Omega}
J(\vec{\Omega}) d \Omega
\label{eq:Jbar}
\end{equation}

The solid angles $\Delta \Omega = 10^{-3} {\rm sr}$ and
$\Delta \Omega = 10^{-5} {\rm sr}$ correspond to the angular resolutions 
in the EGRET experiment and the HESS and GLAST ones, respectively.
The expected gamma-ray flux in an experiment with
threshold energy $E_T$ is finally expressed as
\begin{equation}
 \Phi_{\gamma}(E_T)=1.87 \times 10^{-13} {\rm cm^{-2} s^{-1}} \ \bar{J}(\Delta \Omega) \Delta \Omega
\nonumber
\end{equation}
\begin{equation}
\times \frac{1}{2}
\sum_i \int_{E_T}^{m_{\chi}} d E_{\gamma} \frac{d N^{i}_{\gamma}}{d
  E_{\gamma}} \left( \frac{\langle \sigma_i v \rangle}{10^{-29}{\rm cm^3
      s^{-1}}} \right) \left(\frac{100 {\rm GeV}}{m_{\chi}} \right)^2
\label{flux-int}
\end{equation}

To compute the quantity $\bar{J}$, we assume the following effective 
parameterization for the dark matter halo
\begin{equation}
  \rho(r)= \rho_0 \frac{[1+(R_0/a)^{\alpha}]^{(\beta-\gamma)/\alpha}}
  {[1+(r/a)^{\alpha}]^{(\beta-\gamma)/\alpha}}
  \(\frac{R_0}{r}\)^{\gamma} \;\;
  \(1+\(\frac{r_{sp}}{r}\)^{\gamma _{sp}-\gamma}\) \;\;,
\label{eq:alphabetagamma} 
\end{equation}
\begin{center} 
\begin{table} [t]
\centering 
\begin{tabular}{|c|cccccc|} 
\hline 
&$\alpha$&$\beta$&$\gamma$&a (kpc)&$\bar{J}\left( 10^{-3} {\rm sr} \right)$&$\bar{J}\left( 10^{-5} {\rm sr} \right)$ \\ 
\hline 
Iso& 2.0& 2.0& 0& 3.5&$2.46 \times 10^1$&$2.47 \times 10^1$\\ 
Kra& 2.0& 3.0&0.4 & 10.0 &$ 1.932 \times 10^1$&$2.37 \times 10^1$ \\ 
NFW& 1.0& 3.0& 1.0& 20& $1.21 \times 10^3$& $1.26 \times 10^4$\\ 
Moore& 1.5& 3.0& 1.5& 28.0 &$ 1.60 \times 10^5$&$1.24 \times 10^7$ \\ 
\hline 
\end{tabular} 
\caption{Parameters of some widely used non spiky density profiles models 
and corresponding value of $\bar{J}(10^{-3} {\rm sr})$ and  $\bar{J}(10^{-5} {\rm sr})$.} 
\label{tab} 
\end{table} 
\end{center} 
\begin{figure}
\begin{center}
\begin{tabular}{c}
\psfrag{logjb}[c][ct]{\normalsize $\log_{10}{[\bar{J}(\Delta \Omega)]}$}
\psfrag{sp2}[c][c]{\scriptsize $\gamma_{sp}=2.5$}
\psfrag{sp1}[c][c]{\scriptsize $\gamma_{sp}=2.25$}
\psfrag{nosp}[c][c]{\scriptsize no spike}
\psfrag{mchi100}[c][c]{\scriptsize $m_{\chi}=100 \ {\rm GeV}$}
\psfrag{sigv}[c][b]{\normalsize $\log_{10}{[\langle\sigma v\rangle \ ({\rm cm^3.s^{-1})}]}$}
\includegraphics[width=0.4\textwidth]{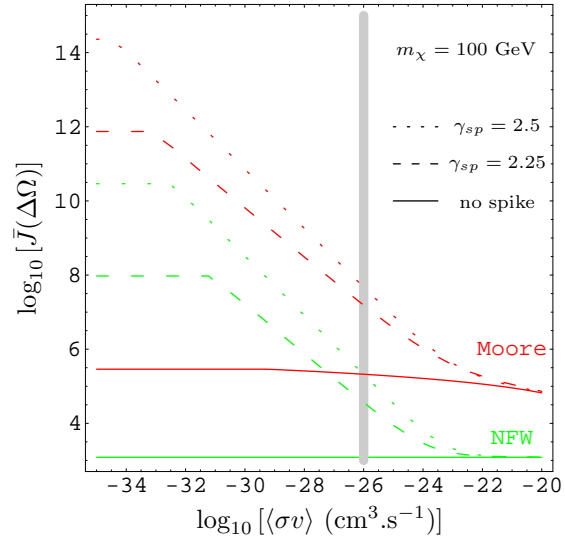}
\end{tabular}
\caption{\small Variation of $\bar{J}(10^{-3} \ {\rm sr})$ with $\langle \sigma
  v \rangle$ for
NFW and Moore profiles, with and without spike. The WMAP 
constraint on relic
density suggests few orders of magnitude around the vertical grey line.}
\label{fig:spike}
\end{center}
\end{figure}
as suggested by many $\Lambda$CDM simulations. 
$R_0$ is the Sun's distance to the galactic center,
$\rho_0$ is the solar neighborhood halo density and $a$ is a characteristic length.
The exponents $\alpha$, $\beta$ and $\gamma$ can be thought of as power law
indices characteristic for $r \simeq a$, $r \gg a$ and $r \ll a$, respectively.
Table~\ref{tab} gives the parameters for common halo models like
the isothermal one which is not cuspy at the center, 
behaving as $\rho(r) \propto {\it cst}$,
and the halo models from $\Lambda$CDM simulations by 
Kravtsov  {\it et al}~\cite{kra}, Navarro, Frenk and White 
(NFW)~\cite{nfw} and Moore {\it et al}~\cite{moore}, which behave
respectively as $\rho(r)\propto r^{-0.4}$, $\rho(r)\propto r^{-1}$ and  
$\rho(r)\propto r^{-1.5}$ at small $r$.
Adiabatic accretion of dark matter on the SBH could further add a central 
spike to these profiles \cite{silkgondo}, or, if more realistic
physics as off-centered formation of the SBH is taken into account,
the SBH will simply enhance the 
cusp \cite{zhao}. Finally, scattering of dark matter 
particles by stars would substantially decrease the density in the center-most
region \cite{gnedin,merrit04}. In the next section we will consider
both the NFW and the Moore  
profiles with and without spike. The spike's characteristic size and slope
are parameterized by $r_{sp}$ and $\gamma _{sp}$ in Eq.~(\ref{eq:alphabetagamma}).
We take (see~\cite{silkgondo}) $r_{sp}=0.35 \, {\rm pc}$, and the two limiting
values for the slope, $\gamma_{sp}=2.25$ and $\gamma_{sp}=2.5$, as $\gamma_{sp}=(9-2\gamma)/(4-\gamma)$.
It is worth emphasizing that the gamma flux is dominated by the contribution
from the {\it unknown} innermost region, especially for profiles with a steep
slope near the center.

\subsection{Annihilation effect on dark matter density}

The astrophysical factor $\bar{J}$ for the annihilation flux becomes formally
divergent for $\gamma \geq 1.5$ ({\it a fortiori} with a spike). 
The presence of a SBH at the center of 
the galaxy solves the problem in principle since no signal will escape  from 
the region inside the Schwarzschild radius $R_S$ ($R_S \sim 3 \cdot 10^{-10} \, {\rm kpc \  for} \  M_{BH}=2.6 \cdot 10^6 M_{\odot}$) .
 The sphere of influence of the SBH is actually larger, as captured particles
 in a region of a few $R_S$ are not balanced by particles scattered in from
 the outer shells. Following~Refs.\cite{gnedin},\cite{merrit04}, we cut the
 density below $r_c=10^{-9}\;{\rm kpc}$: 
\begin{equation}
\rho(r<r_c)=0 
\label{rhocut}
\end{equation}
For halo profiles with a strong cusp/spike behavior near the center,
the density becomes so high that the influence of annihilations on the 
central density has to be taken into account.
A simple bound can be obtained by letting annihilations operate in a static
halo (or with an isotropic velocity distribution~\cite{BM}) initially 
with an infinite density, and during a time as long as the SBH formation
time. By solving the equation 
\begin{equation}
\frac{d n}{dt}=-\langle \sigma v\rangle \; n^2
\end{equation}
one obtains the upper bound
\begin{equation}
\rho_{max}=\frac{m_{\chi}}{\langle \sigma v\rangle t_{BH}} \, .
\end{equation}
In a sense, astrophysics and particle physics aspects can no longer be
decoupled any more in the flux calculation, especially for profiles
with high $\gamma$ values ($ > 1.5$).

\begin{figure}[t]
\begin{center}
\begin{tabular}{c}
\psfrag{a}[c][c]{\small a)}
\psfrag{Jbar}[c][c][1][90]{\small $\log_{10}{[\bar{J}(\Delta \Omega)]}$}
\psfrag{phi}[c][c]{\small $\phi$}
\psfrag{aoverb}[c][c]{\small $a/b$}
\psfrag{1}[c][c]{\small 1}
\psfrag{2}[c][c]{\small 2}
\psfrag{3}[c][c]{\small 3}
\psfrag{4}[c][c]{\small 4}
\psfrag{5}[c][c]{\small 5}
\psfrag{6}[c][c]{\small 6}
\psfrag{8}[c][c]{\small 8}
\psfrag{10}[c][c]{\small 10}
\psfrag{Moore}[c][c]{\normalsize Moore}
\psfrag{NFW}[c][c]{\normalsize NFW}
\psfrag{Kra}[c][c]{\normalsize Kra}
\psfrag{Iso}[c][c]{\normalsize Iso}
\includegraphics[width=0.4\textwidth]{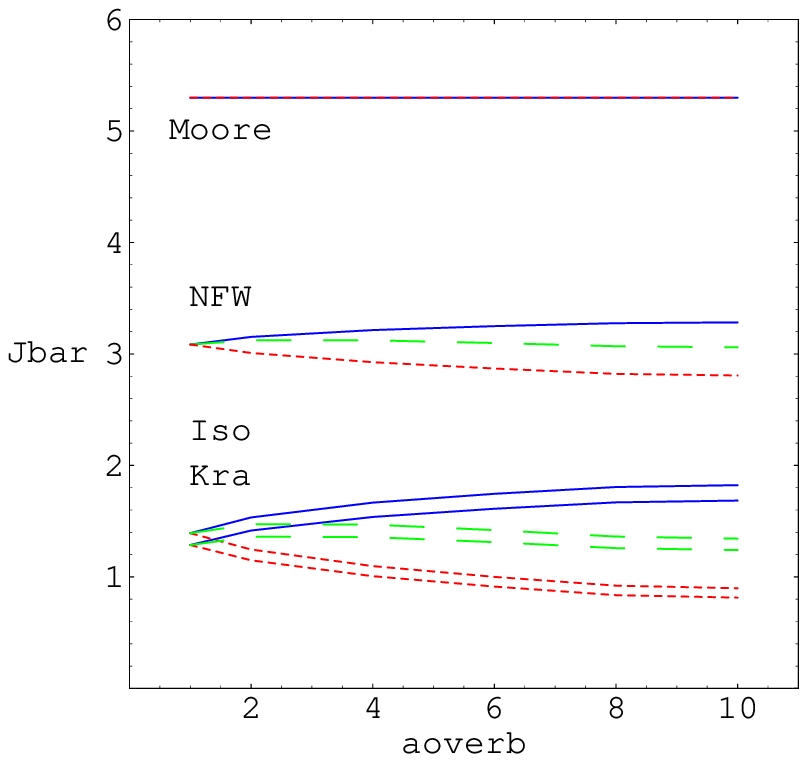}\\
\psfrag{b}[c][c]{\small b)}
\psfrag{Jbar}[c][c][1][90]{\small $\log_{10}{[\bar{J}(\Delta \Omega)]}$}\psfrag{phi}[c][c]{\small $\phi$}
\psfrag{1}[c][c]{\small 1}
\psfrag{2}[c][c]{\small 2}
\psfrag{3}[c][c]{\small 3}
\psfrag{4}[c][c]{\small 4}
\psfrag{5}[c][c]{\small 5}
\psfrag{6}[c][c]{\small 6}
\psfrag{Moore}[c][c]{\normalsize Moore}
\psfrag{NFW}[c][c]{\normalsize NFW}
\psfrag{Kra}[c][c]{\normalsize Kra}
\psfrag{Iso}[c][c]{\normalsize Iso}
\includegraphics[width=0.38\textwidth]{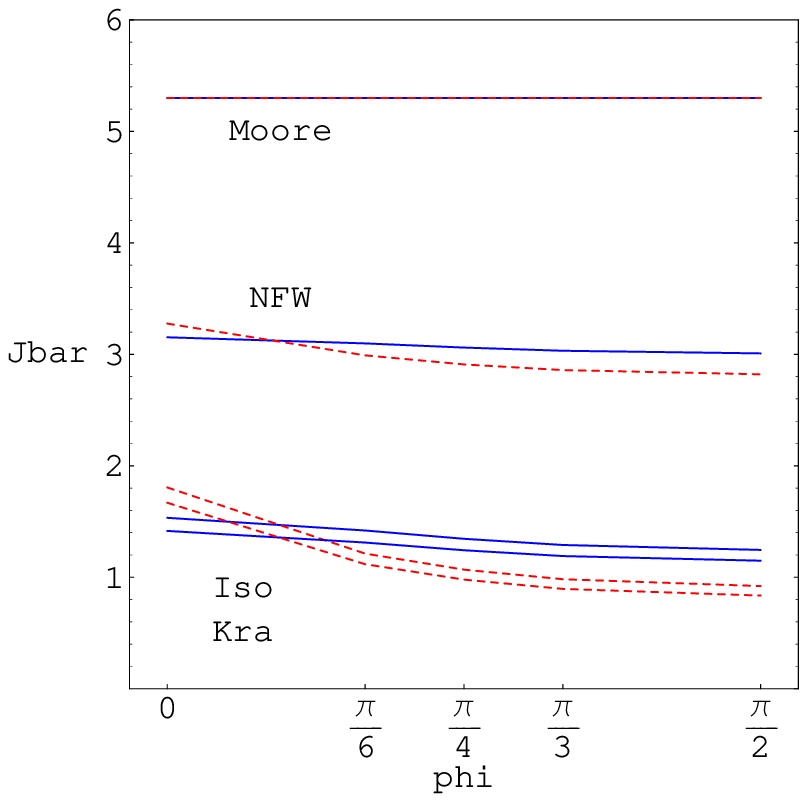}\\
\end{tabular}
\caption{\small $\bar{J}(10^{-3} \ {\rm sr})$ as a function of 
a) the axis ratio $a/b$,  for $\phi=0$ (blue solid line),  $\phi=0.35$
($=20$ deg, green long-dashed line) and $\phi=\pi/2$ (red short-dashed line); 
b) the angle $\phi$, for   $a/b=2 $ (blue solid line) and  $a/b=9 $
(red dashed line).}  
\label{fig:a-phi}
\end{center}
\end{figure}

In Fig.~\ref{fig:spike}, we can see that the presence of a spike near the SBH strongly enhances the annihilation signal.
The precise value of $\bar{J}$ is very sensitive
to $r_{sp}$, $\gamma _{sp}$ and $\langle\sigma v\rangle/m_\chi$. For very small values of
the annihilation cross-section, the cut due to the capture by the SBH becomes apparent.
For large values of the annihilation cross-section, the spike gets washed out 
and we recover the value for the profile without spike.
Additional effects such as scattering of dark matter particles by stars or with baryons 
have been considered in the literature~(\cite{BM}), they compete with the
annihilations for small values of $\langle\sigma v\rangle$, and need to be
included in a precise evaluation of $\bar{J}$.

Our main purpose, however, is to
emphasize that, if spike or strong cusp behaviors are absent, other
geometrical effects such as the asphericity of the halo become crucial in the
 $\bar{J}$ calculation, as will be shown hereafter.

\begin{figure}[h!]
\begin{center}
\begin{tabular}{c}
\psfrag{Jbar}[c][c][1][90]{\small $\log_{10}{[\bar{J}(\Delta \Omega)]}$}
\psfrag{c}[c][c]{\small {\darkblue $c/a$} \ ({\darkred $--\ a/c$})}
\psfrag{1}[c][c]{\small 1}
\psfrag{2}[c][c]{\small 2}
\psfrag{3}[c][c]{\small 3}
\psfrag{4}[c][c]{\small 4}
\psfrag{5}[c][c]{\small 5}
\psfrag{6}[c][c]{\small 6}
\psfrag{8}[c][c]{\small 8}
\psfrag{10}[c][c]{\small 10}
\psfrag{Moore}[c][c]{\normalsize Moore}
\psfrag{NFW}[c][c]{\normalsize NFW}
\psfrag{Kra}[c][c]{\normalsize Kra}
\psfrag{Iso}[c][c]{\normalsize Iso}
\includegraphics[width=0.38\textwidth]{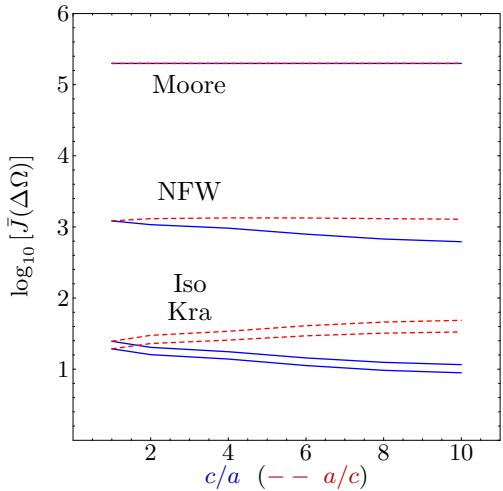}
\end{tabular}
\caption{\small $\bar{J}(10^{-3} \ {\rm sr})$ as a function of prolate
  (blue solid line) and oblate (red dashed line) deformations.}
\label{fig:pro-ob}
\end{center}
\end{figure}

\subsection{Halo asphericity}

A general triaxial halo is modeled by taking 
\be
r=\[(x/a)^2+(y/b)^2+(z/c)^2 \]^{1/2} \, , 
\label{rell}
\ee
with the semi-major axis in the galactic plane aligned with the $x$ axis, $a>b$, and $abc=1$ so that the overall halo mass is fixed.
To study the influence of the non axisymmetry of the halo in the galactic plane, we set $c=1$ and derive the variation
of $\bar{J}$ as a function of $a/b$ and/or the angle $\phi$ between 
the x axis and the direction of the Sun.
Note that an elliptical deformation of the halo can impact
substantially the dark matter annihilation flux without jeopardizing 
the rotation curve fit.
As shown in numerical simulations~\cite{ath1}, the elliptical 
deformation is powered by the angular momentum exchange between the galactic 
bar and the dark halo. When this mechanism is acting solely, the deformation 
is strongest near the galactic center and decreases outwards down to a 
spherical symmetry. In principle, one could use
any radius-dependent axis ratio function $(a/b)(r)$ and recover axisymmetry 
at large  distances 
({\it i.e.} away from the bar).
For simplicity, however, we will consider a constant ellipticity factor. 
This will not introduce a strong deviation from a realistic case because the 
annihilation signal is dominated by the contribution
near the galactic center.
Axial ratios factors of up to 5 have been obtained for the bar in the
baryonic disk component both in observations and in numerical 
simulations, but the oval deformation of the dark halo is expected to be 
milder~\cite{ath2}. Given the uncertainties, however, and the unknowns 
from the history of bar formation that could well vary from one galaxy to 
another, we examine here a wider range of
possible values.

\begin{figure}[t!]
\begin{center}
\begin{tabular}{cc}
\includegraphics[width=0.45\textwidth]{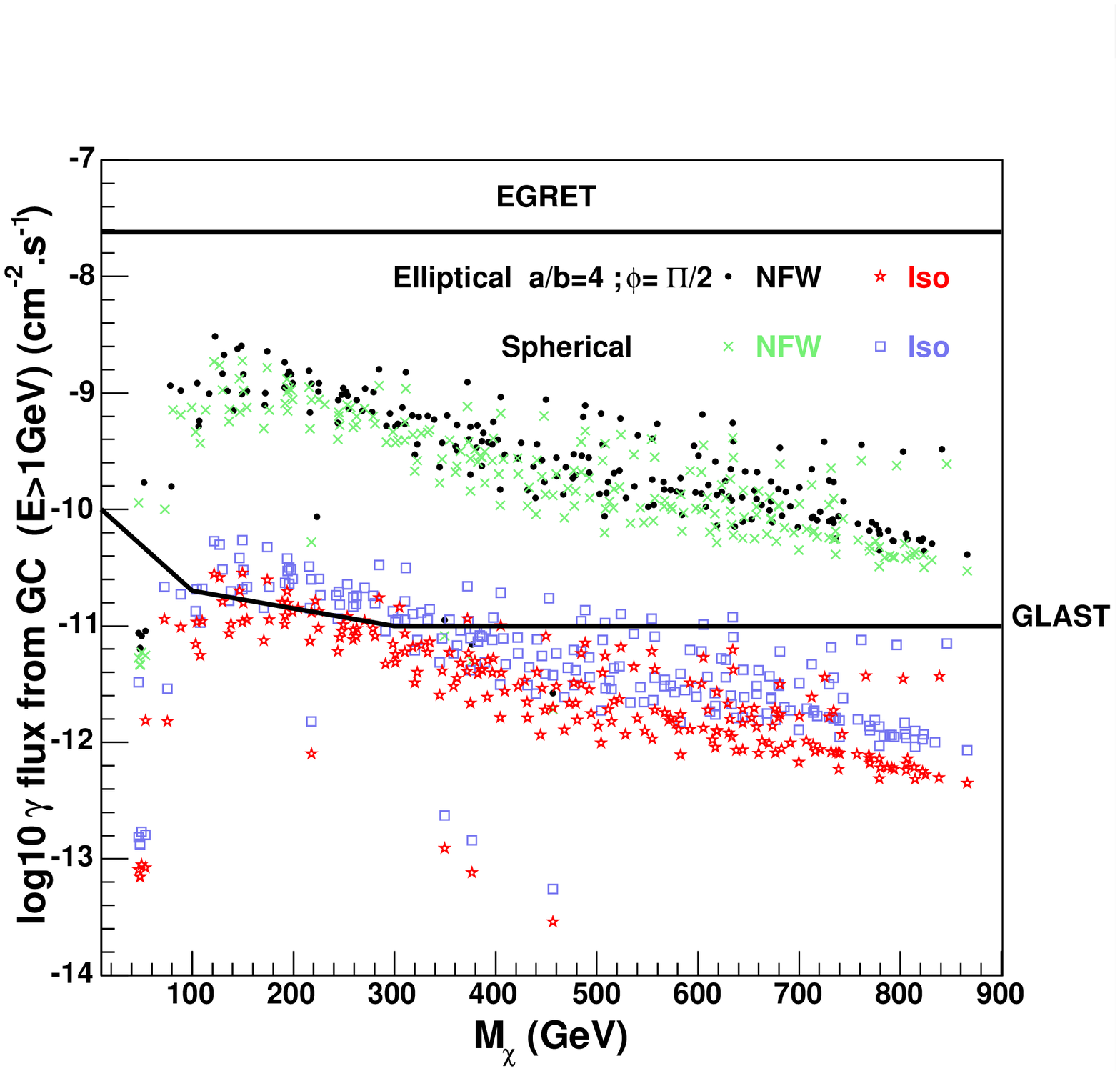}\\
a)\\
\includegraphics[width=0.45\textwidth]{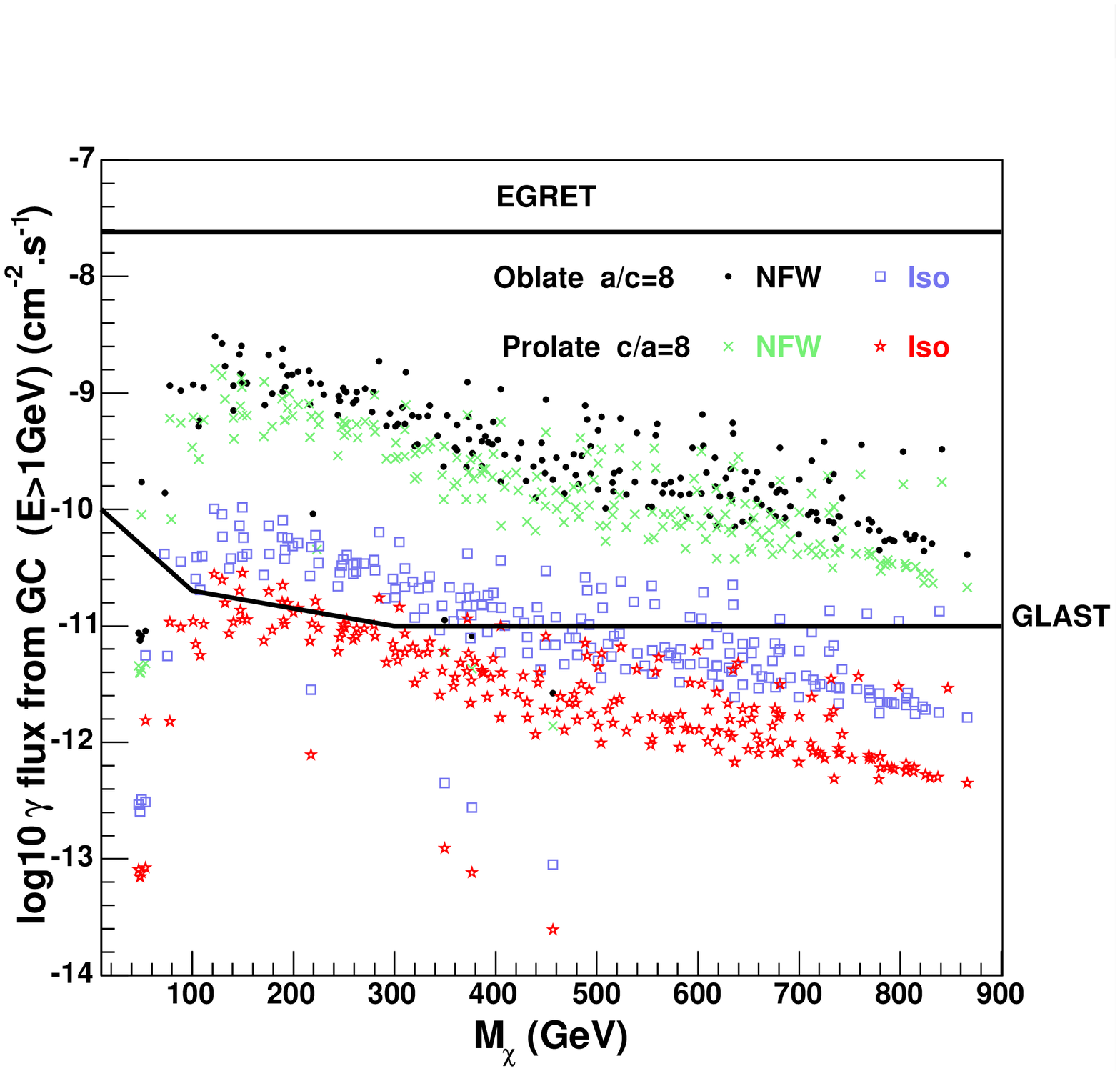}\\
 b)\\
\end{tabular}
\caption{\small Effects on gamma fluxes of NFW and Isothermal halo elliptical
  deformations : 
a) in the disc plane b) prolate and oblate cases .}
\label{fig:flux-exp}
\end{center}
\end{figure}

In Fig.~\ref{fig:a-phi}, the variations of $\bar{J}(\Delta \Omega = 10^{-3}
{\rm sr})$ as a function  of $a/b$ for $\phi=0, \,0.35$ (= 20 deg; see
review \cite{wdehnen} and references therein) and $\pi/2$, and as a
function of $\phi$ 
for $a/b=2$ and 9 are given for the isothermal, Kravtsov, NFW and Moore halo profiles.
For $\phi=0$, $\bar{J}$ increases with $a/b$, which is expected since
the higher density region is stretched along the line of sight in this
case. For larger values of $\phi$, the variation becomes negative,
as the stretching in the x axis is misaligned with the line of sight.
Therefore, $\bar{J}$ is a decreasing function of $\phi$ for a fixed 
value of $a/b$.

It is important to notice that the impact of the ellipticity
is stronger for less cuspy halo profiles (which are favored by
observations). Indeed, the relative contribution  
to $\bar{J}$ coming from inner regions
inside a small radius $r$ increases with $\gamma$. Therefore, for higher
values of $\gamma$, a larger fraction of the volume integral 
$\int \rho^2 dV$ is not affected by a change in ellipticity, as
the observation solid angle is taken constant.

To study the influence of the prolate-oblate shape of the halo,
we set $a=b$ in Eq.~(\ref{rell}) and let $c/a$ vary. The result is
given in Fig.~\ref{fig:pro-ob}, for the isothermal, Kravtsov,
NFW and Moore halo profiles. 
The variation of $\bar{J}$ with the prolate-oblate shape of the halo is again 
stronger for smaller values of $\gamma$.
As we can see on Fig. \ref{fig:pro-ob}, an oblate deformation induces an
enhancement of dark matter density along the line of sight increasing 
the signal, whereas the prolate shape decreases it and thus could be
understood with arguments similar to those above.

Finally, let's consider the most popular dark matter candidate, {\it i.e.} the
  neutralino ($\chi$), which comes from the neutral gauge and Higgs boson
  superpartners in the {\it  Minimal Supersymmetric Standard Model} (MSSM) framework. 
We show in Fig. \ref{fig:flux-exp} the neutralino dark 
matter resulting fluxes for a wide sample of supersymmetric models, {\it i.e} we take parameters of the MSSM to get bino as well as mixed bino-wino and bino-higgsino neutralino 
which have higher couplings and cross sections (see {\it e.g.} \cite{mynonuniv}). All the points shown satisfy
the WMAP requirement on relic density and accelerator constraints. As the GLAST
experiment sensitivity will probe a wide range of halo profiles, we clearly 
see that in addition to the (essentially inner) power law behavior of the 
halo, the geometry also alters the estimation of the fluxes and has to be 
included in flux calculations.  

\section{Conclusions}

The dark matter annihilation signal from the galactic center has been
calculated for different halo characteristics. 
In particular, we have  shown some 
possible effects of the halo asphericity. The induced corrections are more 
relevant for flat than for cuspy cores.

Although a plausible elliptical deformation of the dark matter halo
does not change the expected annihilation signal by orders of magnitude,
a consistent prediction of the flux from the halo shape or conversely 
of the halo shape from the signal should take those effects into account. 

\section*{Acknowledgments}

The authors thank Albert Bosma and Jean Orloff for interesting and 
motivating discussions and O. Gnedin and H. Zhao for useful email exchanges.
F.-S. L. and E.N. work is supported by the I.I.S.N. and the 
Belgian Federal Science Policy (return grant and IAP 5/27).

\end{document}